\documentclass[aps,pra,superscriptaddress,amsmath,amssymb,preprintnumbers,floatfix,showpacs,11pt]{revtex4}

\usepackage{amssymb}
\usepackage{epsfig}

\begin{document}
\title{ Quantum phase properties of two-mode
Jaynes-Cummings model for Schr\"{o}dinger-cat states: interference
and entanglement}

\author{ Faisal A. A. El-Orany }
\affiliation{ Department of Mathematics  and Computer Science,
Faculty of Science, Suez Canal University,
 Ismailia, Egypt}

\author{ M. H. Mahran }
\affiliation{Faculty of Computers $\&$ Informatics, Suez Canal
University,
 Ismailia, Egypt}

\author{ M. R. B. Wahiddin and A. M. Hashim }
\affiliation{ Centre for Computational and Theoretical Sciences,
Kulliyyah of Science, International Islamic University Malaysia,
53100 Kuala Lumpur, Malaysia}

\date{\today}

\begin{abstract}
In this paper we investigate the quantum phase properties for
 the coherent superposition states (Schr\"{o}dinger-cat states)
 for two-mode multiphoton
Jaynes-Cummings model in the framework of the Pegg-Barnett
formalism.
We also demonstrate the behavior of the Wigner ($W$) function at the
phase space origin. We obtain many interesting results such as there is
a clear relationship between the revival-collapse phenomenon occurring in
the atomic inversion (as well as in the evolution of the $W$ function)
and the behavior
of the phase distribution of both the single-mode and two-mode cases. Furthermore,
we find that the phase variances of the single-mode case  can
exhibit revival-collapse phenomenon about the long-time behavior.
We show that such behavior occurs for interaction time several times smaller
than that of the single-mode Jaynes-Cummings model.
\end{abstract}

  \maketitle

\section{Introduction}
One of the few exactly solvable fully quantum-mechanically models describing
the interaction between the single-mode electromagnetic wave
and two-level atom   is the Jaynes-Cummings model JCM \cite{jay1}.
Various nonclassical effects have been reported for this system, in
particular,  when the field is initially prepared in the coherent light.
For instance, the
revival-collapse phenomenon (RCP) in the atomic inversion \cite{eber},
quadrature squeezing \cite{meys} and sub-Poissonian statistics
\cite{kim2}. Furthermore, JCM has been used as a source for generating nonclassical
states in the framework  of the conditional measurement technique \cite{cm1}.
For more details related to JCM reader can consult the review paper, e.g.,
 \cite{bruce}.
Actually, this simple model, i.e. single-mode interacts with two-level
atom, has been extended to include multimode
fields \cite{multimode}, multilevel atoms \cite{multil},
multiatom interactions \cite{multia}, and damping \cite{multid}.
On the other hand, several experiments have been performed
for the JCM, e.g., isolated single atoms \cite{isol}, a single-mode
two-photon Rydberg atom "micromaser" \cite{micro} and observation of
RCP in the evolution of the atomic inversion \cite{remp}.

The progress in the generation of quantum mechanical states, e.g. in
a trapped ion \cite{[39]},
nondemolation measurement \cite{nond}, conditional measurement \cite{cm1},
..etc.,
encourages researchers to analyse the behavior of the JCM  for
these states.  One of these states  is the Schr\"{o}dinger-cat
state \cite{sch1}, which can exhibit various of the nonclassical
effects arising from the interference in phase space.
These  states can be represented (for the $j$th mode of convenience) as

\begin{eqnarray}
\begin{array}{lr}
|\alpha\rangle_{\epsilon_{j}}=N_{j} [|\alpha_{j}\rangle+\epsilon_{j}
|-\alpha_{j}\rangle],\\
\\
=\sum\limits_{n=0}^{\infty}C^{(j)}_{n}|n\rangle,
\label{1}
\end{array}
\end{eqnarray}
where $|\alpha_{j}\rangle$ is a coherent state of the $j$th mode
with complex amplitudes $\alpha_{j}$;
$\epsilon_{j}$ is a c-number takes on values, $0,1,-1$ denoting
coherent states, even coherent states and odd coherent states.
Also  $N_{j}$ is the normalization constant
having the form

\begin{equation}
N^{2}_{j}=\frac{1}{[1+\epsilon_{j}^{2}+2\epsilon_{j}\exp
(-2|\alpha_{j}|^{2})]}.\label{2}
\end{equation}
In the second line of (\ref{1}) we have expressed the states as a
linear combination of Fock states, where the probability distribution amplitudes are
\begin{equation}
C^{(j)}_{n}=N_{j}\exp(-\frac{1}{2}|\alpha_{j}|^{2})
\frac{\alpha_{j}^{n}}{\sqrt{n!}}[1+(-1)^{n}\epsilon_{j}].
\label{2a}
\end{equation}
Throughout this paper we use the shorthand notation $C^{(j)}_{n_{j}}$
instead of $C^{(j)}_{n_{j}}(\epsilon_{j})$. The
two-mode density matrix of the cat state (\ref{1}), which includes $16$
elements, can be briefly written as
\begin{eqnarray}
\begin{array}{rl}
\hat{\rho}_{f}(0)=|\alpha\rangle_{\epsilon}|\alpha\rangle_{\epsilon'}
\mbox{$_{\epsilon'}$}
\langle\alpha|\mbox{$_{\epsilon}$}\langle\alpha|\\
\\
=\hat{\rho}_{I}+\hat{\rho}_{M}, \label{extra}
\end{array}
\end{eqnarray}
where $\hat{\rho}_{M}$ and $\hat{\rho}_{I}$ are the statistical-mixture
and interference parts whose exact forms are given in \cite{faisal2}.
As we mentioned above that for the single-mode cat states
the origin of the occurrence of the nonclassical effects is in
the quantum  interference between the components of the states.
 For instance, coherent states are close to
classical states, however, even (odd) coherent states exhibit
significant squeezing (sub-Poissonian statistics) and their photon
number distributions provide  oscillatory behavior.
It is worth reminding that the evolution of the Schr\"{o}dinger-cat states with
the single-mode JCM has
been treated  in \cite{vid,{ger1},{jos},{faisal1}}.

The phase properties of the JCM  using
the Pegg-Barnett formalism \cite{pegg} have been investigated in, e.g. \cite{phas1,
{phas2},{phas3}}, in
particular, when the field is initially prepared in the coherent state.
It has been shown that for the single-mode JCM
the evolution of both the phase variance and the phase
distribution can carry certain information on the RCP of the corresponding atomic inversion
\cite{phas1}. Moreover, the phase variances of the  multiphoton single-mode
JCM exhibit RCP in the course of the long-time interaction  \cite{phas2}.
Recently, the phase properties of the two-mode JCM (TMJCM) in the form of
linear interaction rather than nonlinear version is studied in \cite{obad1}.
In the present  paper we investigate the phase properties of the TMJCM whose
 Hamiltonian is described by the multiphoton hyper-Raman process and
the modes are initially prepared in the Schr\"{o}dinger-cat states (\ref{extra}).
We use Pegg-Barnett formalism to perform such investigation.
There are various  mechanisms controlling the evolution of the
considered system, e.g.,
interference in phase space, entanglement between the two modes (boson-boson
interaction) as well as  between the two modes
and the two-level atom.
The aim of the present analysis is to investigate
the influence of these mechansims on the behavior of  the phase distribution.
Needless to say that the dynamics of the system
are more complicated than those of the single-mode JCM.
Moreover,  such  investigation is motivated by
 several facts. For instance,
entanglement is connecting with the quantum computers and quantum cryptography
\cite{eker}.
Additionally, decoherence is at the heart of quantum measurement theory \cite{dech}.
Also the phase distribution has been realised experimentally by means of the
 optical homodyne tomography \cite
{tom}, which allows quantum phase mean values to be calculated via the measured
field density matrix.

In this paper we obtain many interesting  results for TMJCM.
For example, there is a good connection between the behavior of the atomic inversion
and that of the corresponding phase distribution of both the single-mode
and compound-mode cases.
 Also we show that under certain condition there
is a relationship between the evolution of the atomic inversion and the
 $W$ function at the phase space origin.
Furthermore, the evolution of
 the phase variance of the single-mode case  can
exhibit RCP about the long-time behavior, which occurs on the time domain
 several times smaller
than that of the single-mode JCM.
We analyse the behavior of the TMJCM in the following  order: In section 2 we give the basic
equations and relations, which will be used in the paper.
Section 3 is devoted to discuss the main results.
Summary of the results is given in section 4.
\section{Basic equations and relations}
In this section we describe the system and provide the main
relations used in the paper. Such relations include
the Hamiltonian of the system,  the corresponding wave function and
the calculations associated with the Pegg-Barnett formalism for the system
  \cite{pegg}.

We start with the Hamiltonian, which describes the interaction of the
two modes with the single two-level atom and can briefly be written as
\begin{equation}
\frac{\hat{H}}{\hbar}=\hat{H}_{f}+\hat{H}_{I},
\label{6}
\end{equation}
where $\hat{H}_{f}$ and $\hat{H}_{I}$ are the free part and the interaction
part, respectively. In the rotating wave approximation these quantities
take the forms

\begin{eqnarray}
\begin{array}{lr}
\hat{H}_{f}=
\omega_{1}\hat{a}_{1}^{\dagger}\hat{a}_{1}+
\omega_{2}\hat{a}_{2}^{\dagger}\hat{a}_{2}+
\frac{1}{2}\omega_{a}\hat{\sigma}_{z},\\
\\
\hat{H}_{I}=
g(\hat{a}_{1}^{\dagger k_{1}}\hat{a}_{2}^{k_{2}}\hat{\sigma}_{+} +
\hat{a}_{2}^{\dagger k_{2}}\hat{a}_{1}^{k_{1}}\hat{\sigma}_{-}),
\label{6a}
\end{array}
\end{eqnarray}
where $\hat{\sigma}_{\pm}$ and $\hat{\sigma}_{z}$ are the Pauli spin
operators;
$\omega_{j}, (j=1,2)$ and $\omega_{a}$ are the frequencies of
cavity modes $\hat{a}_{j}$ and the atomic transitions, respectively;  $g$ is the atom-field coupling
constant and $k_{j}$ are the transition parameters, i.e. they represent
the number of photons involved in the atomic transition.
The derivation of the Hamiltonian (\ref{6a}) from the first principle is given in
\cite{avia}. Moreover,
(\ref{6a})  generalizes various models given in the literature,
e.g. \cite{chr,{card}}, and for $k_{1}=k_{2}=1$ it can be interpreted as a
 cavity version of
Raman scattering in which a mode $1$ is the pump field,  mode $2$ is
the Stokes field and with no anti-Stokes field \cite{chr}. In
this case the system represents three-level atom in $\Lambda$
configuration interacting with the two modes of the field under the
assumption of the exact  resonance. Applying adiabatic elimination of
the second level the effective Hamiltonian reduces to the form of the usual
JCM Hamiltonian but the single-mode field operator is replaced by the product of the
 two modes operators.

We proceed by considering the exact resonance case $\omega_{a}=k_{1}\omega_{1}
-k_{2}\omega_{2}$. The dynamical properties of the system
 can be obtained by means of the unitary operator of the interaction part as
\begin{eqnarray}
\begin{array}{rl}
\hat{U}_{I}(T,0)=\exp(-i\frac{T}{g}\hat{H}_{I})
\\
\\
=\cos (T\hat{D})-i\frac{\sin (T\hat{D})}{g\hat{D}}\hat{H}_{I},\label{8}
\end{array}
\end{eqnarray}
where
\begin{equation}
T=g t,\qquad
\hat{D}^{2}=
\hat{a}_{1}^{\dagger k_{1}}\hat{a}_{1}^{k_{1}}
\hat{a}_{2}^{ k_{2}}\hat{a}_{2}^{ \dagger k_{2}}
\hat{\sigma}_{+} \hat{\sigma}_{-}+
\hat{a}_{1}^{ k_{1}}\hat{a}_{1}^{\dagger k_{1}}
\hat{a}_{2}^{ \dagger k_{2}}\hat{a}_{2}^{ k_{2}}
\hat{\sigma}_{-} \hat{\sigma}_{+}. \label{9}
\end{equation}
It is worth mentioning that without the rotating wave approximation,
i.e.  involving the fastly rotating terms in the Hamiltonian, the fully
quantum-mechanical system cannot be solved as
 in (\ref{8}) since the eigenstates of the Hamiltonian cannot be obtained
in a closed form. In this case, numerical techniques such as
path-integral approach \cite{Zah}
 and perterbative approach
\cite{Pho} have to be used. Inspite of  these techniques are very
sophisticated  the obtained solution will be in an approximate form.
Of course, in this case the results of the present paper are no longer valid.
For instance, we can mention for
the single-mode JCM without RWA the atomic inversion is phase dependent
\cite{Pho}.

Additionally, we assume that the two modes are initially prepared in the cat state
(\ref{extra}) and   the atom  is  in the
 superposition of the excited and ground states having the from
\begin{equation}
|\varphi,\phi\rangle=\cos\varphi |+\rangle+\exp(i\phi)\sin\varphi
|-\rangle,\label{2}
\end{equation}
where
$|+\rangle$ and
$|-\rangle$ denote excited and ground atomic states, respectively,
and
$\phi, \varphi$ are the relative phases between $|-\rangle$
and $|+\rangle$.
Therefore the total initial state of the system is
\begin{equation}
|\Psi (0)\rangle=|\alpha\rangle_{\epsilon _1}\bigotimes
|\alpha\rangle_{\epsilon_ 2}\bigotimes
 |\varphi,\phi\rangle. \label{3}
\end{equation}
From (\ref{8})--(\ref{3}) the dynamical state of the system is

\begin{eqnarray}
\begin{array}{lr}
|\Psi (T)\rangle=\hat{U}_{I}(T,0)|\Psi (0)\rangle\\
\\
=\sum\limits_{n,m=0}^{\infty}
\left[ F_{1}(n,m,T)|+,n,m+k_{2}\rangle
+F_{2}(n,m,T)|-,n+k_{1},m\rangle\right],
 \label{8aa}
\end{array}
\end{eqnarray}
where the time-dependent  coefficients have the forms:
\begin{eqnarray}
\begin{array}{lr} F_{1}(n,m,T)=C_{n,m+k_{2}}\cos\varphi
\cos(T\Lambda_{n,m}) -iC_{n+k_{1},m} \exp(i\phi)\sin\varphi \sin(T
\Lambda_{n,m}),
\\
\\
F_{2}(n,m,T)=C_{n+k_{1},m}\sin\varphi \cos(T \Lambda_{n,m})
\exp(i\phi)
-iC_{n,m+k_{2}}\cos\varphi \sin(T\Lambda_{n,m}),
\label{8a}
\end{array}
\end{eqnarray}
whereas
\begin{equation}
\Lambda_{n,m}=\sqrt{\frac{(m+k_{2})!(n+k_{1})!}{n!m!}},
\label{8b}
\end{equation}
and  $C_{n,m}=C^{(1)}_{n}C^{(2)}_{m}$, $C^{(j)}_{n}$ are given by
(\ref{2a}).
From (\ref{8aa})  the entanglement between different components of the
system is readily apparent.

The atomic inversion of the system can be expressed as
\begin{eqnarray}
\begin{array}{lr}
\langle\hat{\sigma}_{z}(T)\rangle=\sum\limits_{n,m}^{\infty}
 \Bigl\{[C^{2}_{n,m+k_{2}}
\cos^{2}\varphi -C^{2}_{n+k_{1},m} \sin^{2}\varphi] \cos(2T\Lambda_{n,m})
\\
\\
+\sin\phi\sin (2\varphi)
C_{n,m+k_{2}}C_{n+k_{1},m}\sin(2T\Lambda_{n,m})\Bigr\}
. \label{10a}
\end{array}
\end{eqnarray}

To calculate the joint phase distribution for the field we find the
reduced density operator for the radiation field $\hat{\rho}_{f}(T)$
 by tracing (\ref{8aa})   over the atomic states as

\begin{eqnarray}
\begin{array}{lr}
\hat{\rho}_{f}(T)={\rm Tr}_{a}[\hat{\rho}(T)]\\
\\
=\sum\limits_{n,m,n',m'}^{\infty}
\Bigl[F_{1}(n,m,T)
F^{*}_{1}(n',m',T)|n,m+k_{2}\rangle
\langle n',m'+k_{2}|\\
\\
+ F_{2}(n,m,T)
F^{*}_{2}(n',m',T)|n+k_{1},m\rangle
\langle n'+k_{1},m'|\Bigr]
.\label{9}
\end{array}
\end{eqnarray}
In the first line of  (\ref{9}) the notation ${\rm Tr}_{a}$
means that we  evaluate the trace over the atomic states.

Now we give briefly  the  relations  of the Pegg-Barnett formalism
\cite{pegg}, which will be used throughout  the paper, however, more
 details about the technique with various applications can be found in \cite{luks}.
This formalism is based on introducing a finite $(s+1)$-dimensional space
$\Psi$ spanned by the number states $|0\rangle,|1\rangle,...,|s\rangle$.
The expectation values of the different hermitian phase operators
can be  evaluated in the finite dimensional space $\Psi$ and  at
the final stage the limit $s\rightarrow \infty$ is taken.
The hermitian phase operator of the single-mode case is
defined as
\begin{equation}
\hat{\Phi}=\sum^{s}\limits_{m=0}\Theta_{m} |\Theta_{m}\rangle
\langle \Theta_{m}|,
\label{8p}
\end{equation}
where  the states $|\Theta_{m}\rangle$ are eigenstates of the phase
operator (\ref{8p}) and they form
a complete orthonormal basis of $s+1$ states in $\Psi$.
Moreover, these states are  restricted to lie within a phase window
between $\Theta_{0}$ and $2\pi+\Theta_{0}$.
For the system described by the density matrix
$\hat{\rho}=\sum^{\infty}\limits_{m,m'=0}C_{m}C^{*}_{m'}|m\rangle \langle m'|$
 the continuum (, i.e. $s$ tends to infinity)
 phase distribution  is defined as
\begin{eqnarray}
\begin{array}{rl}
P(\Theta)={\rm lim}_{s\rightarrow \infty}\frac{s+1}{2\pi}\langle
\Theta_{m}|\hat{\rho}|\Theta_{m}\rangle \\
\\
=\frac{1}{2\pi} \sum\limits_{m,m^{'}=0}^{\infty} C_{m}C^{*}_{m^{'}}
\exp[i(m-m^{'})\Theta],
\end{array} \label{10}
\end{eqnarray}
where $\Theta_{m}$ has been replaced by the continuous phase variable
$\Theta$. Once the phase distribution $P(\Theta)$ is obtained, all
the quantum-mechanical phase moments can be calculated as a classical
integral  over $\Theta$ through the relation
\begin{equation}
\langle
\hat{\Phi}^{l'}\rangle=\int_{-\pi}^{\pi}\Theta^{l'}P(\Theta,T)d\Theta,\qquad
l'=1,2,...
\label{14}
\end{equation}
The phase variance is defined as
\begin{equation}
\langle(\triangle\hat{\Phi})^{2}\rangle=
\langle\hat{\Phi}^{2}\rangle-\langle \hat{\Phi}\rangle^{2}.\label{15b}
\end{equation}

Actually, the generalization of the relations (\ref{8p})--(\ref{15b}) to
the two-mode version  is straightforward.
 In this respect the joint
 phase  distribution associated with the reduced density matrix
 (\ref{9}) can be evaluated  as
\begin{equation}
P(\Theta_{1},\Theta_{2},T)=\frac{1}{4\pi^{2}}
\sum\limits_{n,m,n',m'}^{\infty} C(n,m,n',m',T)
\exp[i(n'-n)\Theta_{1}+i(m'-m)\Theta_{2}] , \label{11}
\end{equation}
where
\begin{eqnarray}
\begin{array}{lr} C(n,m,n',m',T)= F_{1}(n,m,T)F^{*}_{1}(n',m',T)
+F_{2}(n,m,T)F^{*}_{2}(n',m',T)\\
\\
=
[C_{n,m+k_{2}}C_{n',m'+k_{2}}\cos^{2}\varphi
+C_{n+k_{1},m}C_{n'+k_{1},m'}\sin^{2}\varphi]
 \cos[T( \Lambda_{n,m}-\Lambda_{n',m'})].
 \label{10}
\end{array}
\end{eqnarray}
For the sake of simplicity in (\ref{10}) we
restrict the calculation to the case  $\phi=0$ .
Also  one can easily prove that
\begin{equation}
\int_{-\pi}^{\pi}\int_{-\pi}^{\pi}
P(\Theta_{1},\Theta_{2},T)
d\Theta_{1}d\Theta_{2}=1. \label{12}
\end{equation}
The single-mode phase distribution of the $j$th mode
can be obtained from (\ref{11}) via the relation
\begin{equation}
\int_{-\pi}^{\pi}
P(\Theta_{j'},\Theta_{j},T)
d\Theta_{j'}=
P(\Theta_{j},T)
, \quad j\neq j'.\label{13}
\end{equation}
It is worth mentioning that  (\ref{11}) can be expressed in terms of the single-mode phase
distribution as

\begin{eqnarray}
\begin{array}{rl}
P(\Theta_{1},\Theta_{2},T)=\frac{1}{4\pi^{2}}\Bigl\{2\pi\left[P(\Theta_{1},T)+P(\Theta_{2},T)\right]-1\\
\\
+2\sum\limits_{n>n'}^{\infty}\sum\limits_{m>m'}^{\infty}
\Bigl[C_{+}(n,m,n',m',T)\cos [(n-n')\Theta_{1}]\cos[(m-m')\Theta_{2}]\\
\\
+C_{-}(n,m,n',m',T)\sin [(n-n')\Theta_{1}]\sin[(m-m')\Theta_{2}]\Bigr]
\Bigr\}
, \label{11f}
\end{array}
\end{eqnarray}
where
\begin{equation}
C_{\pm}(n,m,n',m',T)=C(n',m,n,m',T)\pm C(n,m,n',m',T).\label{11ff}
\end{equation}
Expression (\ref{11f}) is relevant for numerical tasks.

To investigate the fluctuation in the phase distribution for the system
under consideration we
have to evaluate different moments for the phase operators, which give
\begin{eqnarray}
\begin{array}{lr}
\langle
\hat{\Phi}_{1}\rangle=0,\\
\\
\langle
\hat{\Phi}^{2}_{1}\rangle=
\frac{\pi^{2}}{3}+2\sum\limits_{n\neq n',m}^{\infty}
C(n,m,n',m,T)\frac{(-1)^{n'-n}}{(n'-n)^{2}},\\
\\
\langle\hat{\Phi}_{1}\hat{\Phi}_{2}\rangle=
-\sum\limits_{n\neq n',m\neq m'}^{\infty}
C(n,m,n',m',T)\frac{(-1)^{n'+m'-m-n}}{(m'-m)(n'-n)}.
\label{15a}
\end{array}
\end{eqnarray}
 The sum- and difference-phase operators  are defined as

\begin{equation}
\hat{\Phi}_{\pm}=\hat{\Phi}_{1}\pm\hat{\Phi}_{2}, \label{Meda1}
\end{equation}
where "+" and "-" denote sum-phase and difference-phase operators,
respectively.
Therefore,
the sum- and difference-phase variances
 can be expressed  in the following formula
\begin{eqnarray}
\begin{array}{lr}
\langle(\triangle\hat{\Phi}_{\pm})^{2}\rangle
=\langle(\triangle\hat{\Phi}_{1})^{2}\rangle+
\langle(\triangle\hat{\Phi}_{2})^{2}\rangle\pm h_{1,2},
\\
  \\
h_{1,2}=2[\langle\hat{\Phi}_{1}\hat{\Phi}_{2}\rangle-
\langle \hat{\Phi}_{1}\rangle\langle \hat{\Phi}_{2}\rangle].
\label{15c}
\end{array}
\end{eqnarray}
Moreover, as the mean-photon number and the phase are conjugate
quantities. It is reasonable  to investigate
 the fluctuation in the mean-photon number for  the single-mode and compound-mode cases
$\langle(\triangle\hat{n}_{j})^{2}\rangle$ and
$\langle(\triangle\hat{n}_{\pm})^{2}\rangle$,
 where $\hat{n}_{j}=\hat{a}_{j}^{\dagger}\hat{a}_{j}$ and
 $\hat{n}_{\pm}=\hat{n}_{1}\pm\hat{n}_{2}$.

On the other hand, we address the  relationship between the behaviors of
the  phase distribution and the  $W$ function at the phase space origin for the system under
consideration. The reasons for this will be
 clear shortly.
The joint Wigner function can be easily evaluated using a technique similar
to that given in \cite{faisal1} as
\begin{eqnarray}
\begin{array}{lr}
W(\chi_{1},\chi_{2},T)=\frac{\exp(-|\chi_{1}|^{2}-|\chi_{2}|^{2})}
{\pi^{2}} \sum\limits_{n,m,n',m'}^{\infty}
(-1)^{n'+m'}\chi_{1}^{n-n'}\chi_{2}^{m-m'}2^{\frac{n+m-n'-m'}{2}}\\
\\
\times \Bigl[
(-1)^{k_{2}}F_{1}(n,m,T)
F^{*}_{1}(n',m',T)\sqrt{\frac{n'!(m'+k_{2})!}{n!(m+k_{2})!}}
{\rm L}_{n'}^{n-n'}(2|\chi_{1}|^{2})
{\rm L}_{m'+k_{2}}^{m-m'}(2|\chi_{2}|^{2})
\\
\\
+
(-1)^{k_{1}}F_{2}(n,m,T)
F^{*}_{2}(n',m',T)\sqrt{\frac{(n'+k_{1})!m'!}{(n+k_{1})!m!}}
{\rm L}_{n'+k_{1}}^{n-n'}(2|\chi_{1}|^{2})
{\rm L}_{m'}^{m-m'}(2|\chi_{2}|^{2})\Bigr]
,\label{17}
\end{array}
\end{eqnarray}
where $\chi_{j}=x_{j}+iy_{j}, j=1,2$ and ${\rm L}_{n}^{\nu}(.)$ are
associated Laguerre polynomials of order $n$. The $W$ function of the
single-mode case  can be obtained from (\ref{17})
by tracing over the other mode.
Now the $W$ functions associated with the first mode $(W_{1})$, second mode
 $(W_{2})$ and compound modes $(W)$ at the phase space origin, i.e. $\chi_{
 j}=0$, can be expressed as

\begin{equation}
W_{1}(0,T)=\frac{1}
{\pi}
\sum\limits_{n,m=0}^{\infty} (-1)^{n}
\left[
|F_{1}(n,m,T)|^{2}+ (-1)^{k_{1}}|F_{2}(n,m,T)|^{2}
\right]
,\label{19a}
\end{equation}

\begin{equation}
W_{2}(0,T)=\frac{1}
{\pi}
\sum\limits_{n,m=0}^{\infty} (-1)^{m}
\left[
(-1)^{k_{2}}|F_{1}(n,m,T)|^{2}+ |F_{2}(n,m,T)|^{2}
\right]
,\label{19b}
\end{equation}

\begin{equation}
W(0,T)=\frac{1}
{\pi^{2}}
\sum\limits_{n,m=0}^{\infty} (-1)^{m+n}
\left[
(-1)^{k_{2}}|F_{1}(n,m,T)|^{2}+ (-1)^{k_{1}}|F_{2}(n,m,T)|^{2}
\right]
.\label{19c}
\end{equation}

\begin{figure}
  \includegraphics[width=.9\linewidth]{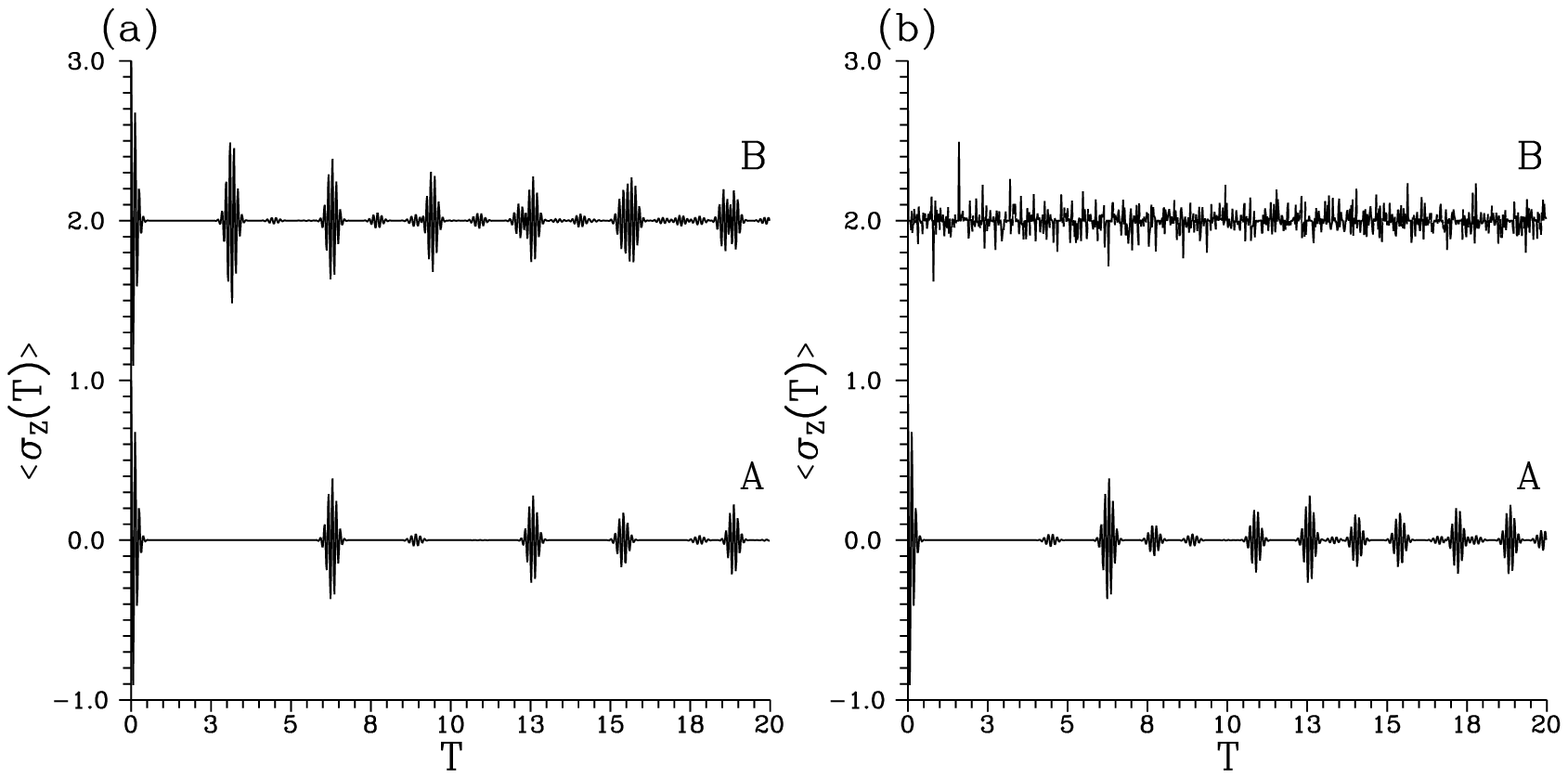}
\caption{ The atomic inversion
$\langle\hat{\sigma}_{z}(T)\rangle$ versus the scaled time $T$ for
$\varphi=0$ and the modes are prepared initially in different
types  of the initial states with $|\alpha_{j}|=5$. (a)
$(\epsilon_{1},\epsilon_{2},k_{1},k_{2})=(0,0,1,1)$ (curve A),
$(1,1,1,1)$ (curve B); (b)
$(\epsilon_{1},\epsilon_{2},k_{1},k_{2})=(1,0,1,1)$ (curve A),
$(1,1,2,2)$ (curve B). The curves B are given for
$\langle\hat{\sigma}_{z}(T)\rangle+2$. }
\end{figure}

Finally, as it is obvious that the system includes various parameters, which
make the analysis to be  difficult. Thus
we restrict ourselves to the parameters, which can give
significant results.  Also in the text the statement
standard JCM  means $k_1=0,k_2=1$, the optical cavity field and atom are
initially prepared
 in coherent state and  in the atomic excited state, respectively.

\section{ Results and discussions}
In this section we investigate the quantities, which have been calculated
in section 2. Such investigation  includes the atomic inversion, $W$
function, phase distribution and and the phase and mean-photon variances.

\subsection{Atomic inversion and Wigner function}
As it is well known for the standard JCM  that there is a direct connection
between the phase distribution and $\langle\hat{\sigma}_{z}(T)\rangle$
\cite{phas1}.
Additionally, quite recently it has been proved that there is a relation
between  $\langle\hat{\sigma}_{z}(T)\rangle$ and the evolution of the
 $W$ function at the phase space origin \cite{faisal1}.
This relationship is supported by the recent developments in photon
counting experiment \cite{con} and trapped ion technique \cite{ion1,{ion2}},
where the
 measurements have been focused on the  origin of the $W$ function
 in the phase space.
The natural question is: what would be the situation for the TMJCM?
In this part we  answer this question by investigating
 the behavior of the
$\langle\hat{\sigma}_{z}(T)\rangle$ and showing its connection with the
$W$ function for the system
under consideration.

Firstly one of the most interesting phenomenon, which is representative to
the single-mode JCM, is the RCP in the evolution of the
$\langle\hat{\sigma}_{z}(T)\rangle$. Much attention has been focused on
this phenomenon because it provides evidence of the granularity of the
radiation field.
It is worth mentioning that  the observation of RCP has been performed
via one-atom
mazer \cite{remp}. Moreover, schemes for measuring RCP via homodyne
detection \cite{faisal4}, photon counting experiment and homodyne
tomography \cite{faisal5} have been reported, too.
 For more details related to such phenomenon
reader can consult, e.g.  \cite{eber}.
On the other hand,
RCP associated with the TMJCM is  very complicated compared to that one
associated with the
standard JCM where for the TMJCM the positions of the
revivals in the time domain are independent of the
intensities of the initial modes provided that
$\langle\hat{n}_{j}(0)\rangle=\bar{n}_{j}>>1$. Such behavior  has been
explained partially in \cite{chr}. Also
 in \cite{card} it has been shown that there are secondary revivals coming
from the complicated interferences, which are resulting from the double
summations in the atomic inversion (cf. (\ref{10a})).
We should stress that the
origin of the RCP in the $\langle\hat{\sigma}_{z}(T)\rangle$ of the TMJCM is
the strong entanglement between
the bosonic and fermonic systems. To be more specific,
restricting the discussion to the case $k_{1}=k_{2}=1$,
when  the atom is classically treated the Hamiltonian (\ref{6}) describes the up
conversion process \cite{in9,{in10}}, which is representative well by  switching
the energy between the signal and idler modes. Conversely,
when  the field is classically treated, i.e. $\hat{a}_{j}\rightarrow
|\alpha_{j}|\exp(-i\bar{\phi}_{j})$, (\ref{6}) reduces to two-level
atom interacting with the classical fields. In this case
$\langle\hat{\sigma}_{z}(T)\rangle$ exhibits steady-state Rabi frequency.
In Figs. 1(a) and (b) we have plotted $\langle\hat{\sigma}_{z}(T)\rangle$
against the scaled time $T$ when the atom is in the excited state and the
modes are initially  prepared  in  different types of
states as indicated.
\begin{figure}
   \includegraphics[width=.90\linewidth]{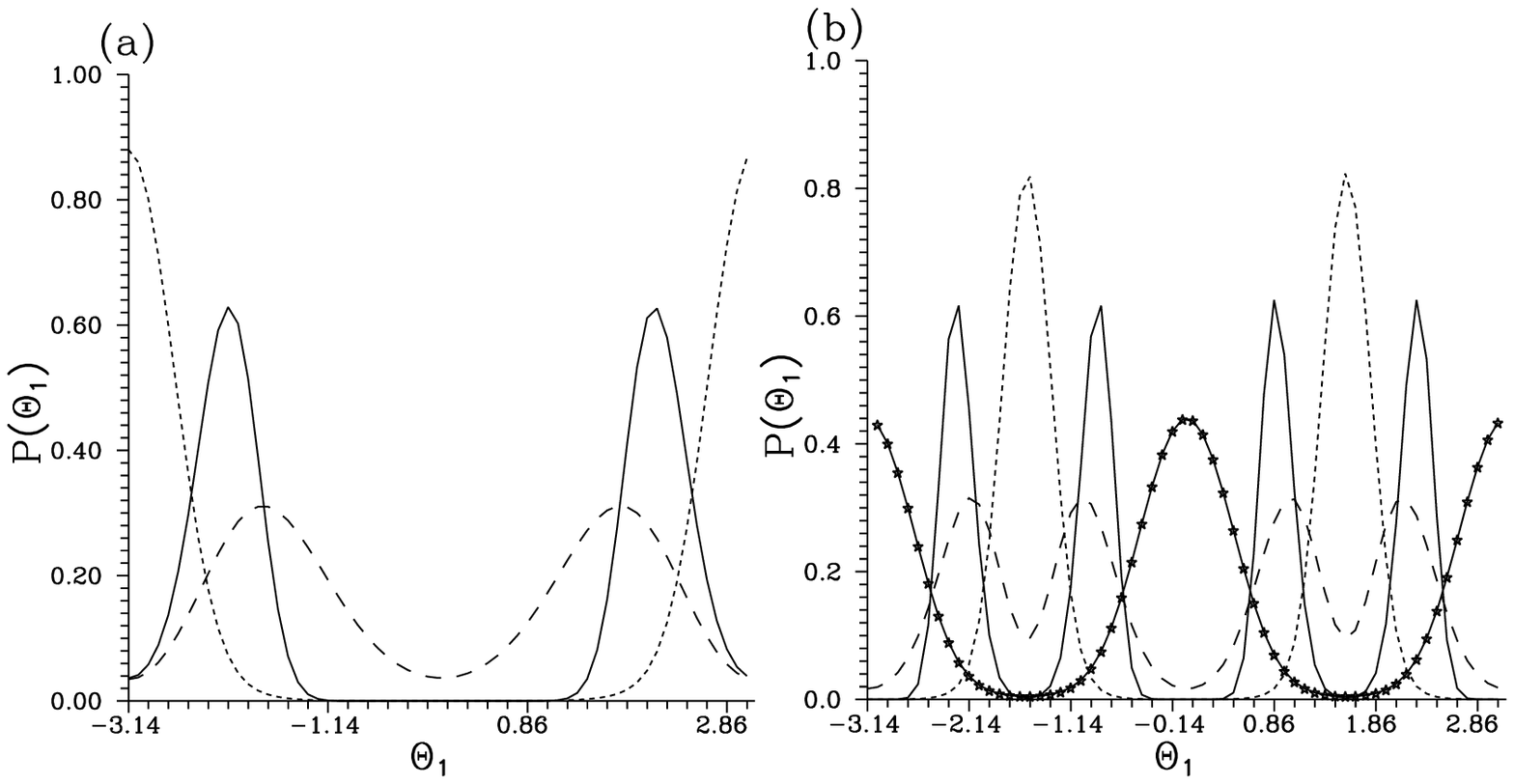}
\caption{ The phase distribution $P(\Theta_{1})$  of the first
mode when $|\alpha_{j}|=5, \varphi=0,\phi=0,k_{1}=k_{2}=1$ and for
$(\epsilon_{1},\epsilon_{2})=(0,0) (a)$
  and  $(1,1) (b)$.
For (a) $T=4.42$ (solid curve-collapse time), $6.2999$ (short-dashed curve--
revival time) and $9.32$ (long-dashed curve--secondary revival time).
For (b) $T=1.8$ (solid curve-collapse time), $3.22$ (short-dashed
 curve--revival time) and $4.4$ (long-dashed curve--secondary revival time).
In (b) the star-centered curve is given for
$(\epsilon_{1},\epsilon_{2})=(0,1)$ and $T=6.3$ (revival time).}
\end{figure}
In Fig. 1(a) for curve A~one can observe  the revivals and secondary
revivals representative to TMJCM.  For curve B,
where the interference in phase space starts to play a role, the
revivals and secondary revivals
 are more pronounced than those shown in
curve A and the behavior becomes more systematic,
i.e. each revival is followed by a secondary revival.
 This can be explained as follows. For curve B the
density matrix (\ref{extra}) has two parts, which are $\hat{\rho}_{M}$ and
$\hat{\rho}_{I}$ (cf. (\ref{extra})). Each of which provides its own RCP pattern
and consequently they interfere with
each other showing such behavior. Furthermore, it seems  that
the revival  time for the curve $A$ is two
times greater than that for curve $B$. This can be partially explained
for $\bar{n}_{j}>>1$ and $k_{1}=k_{2}=1$. In this case the photon distribution for both initial
coherent state and even coherent state possess  Poissonian envelope.
This means that the
terms contributing effectively to the summations in (\ref{10a})
are those for which the pairs $(\alpha_{1},\alpha_{2})$ and $(n,m)$ are comparable.
Now we assume that
the arguments of $\cos (.)$ for initial
coherent light is $\lambda_{1}$ with interaction time $T$ and  for initial
even coherent light is
$\lambda_{2}$ with interaction time $T'$, i.e.
\begin{figure}
 \includegraphics[width=.90\linewidth]{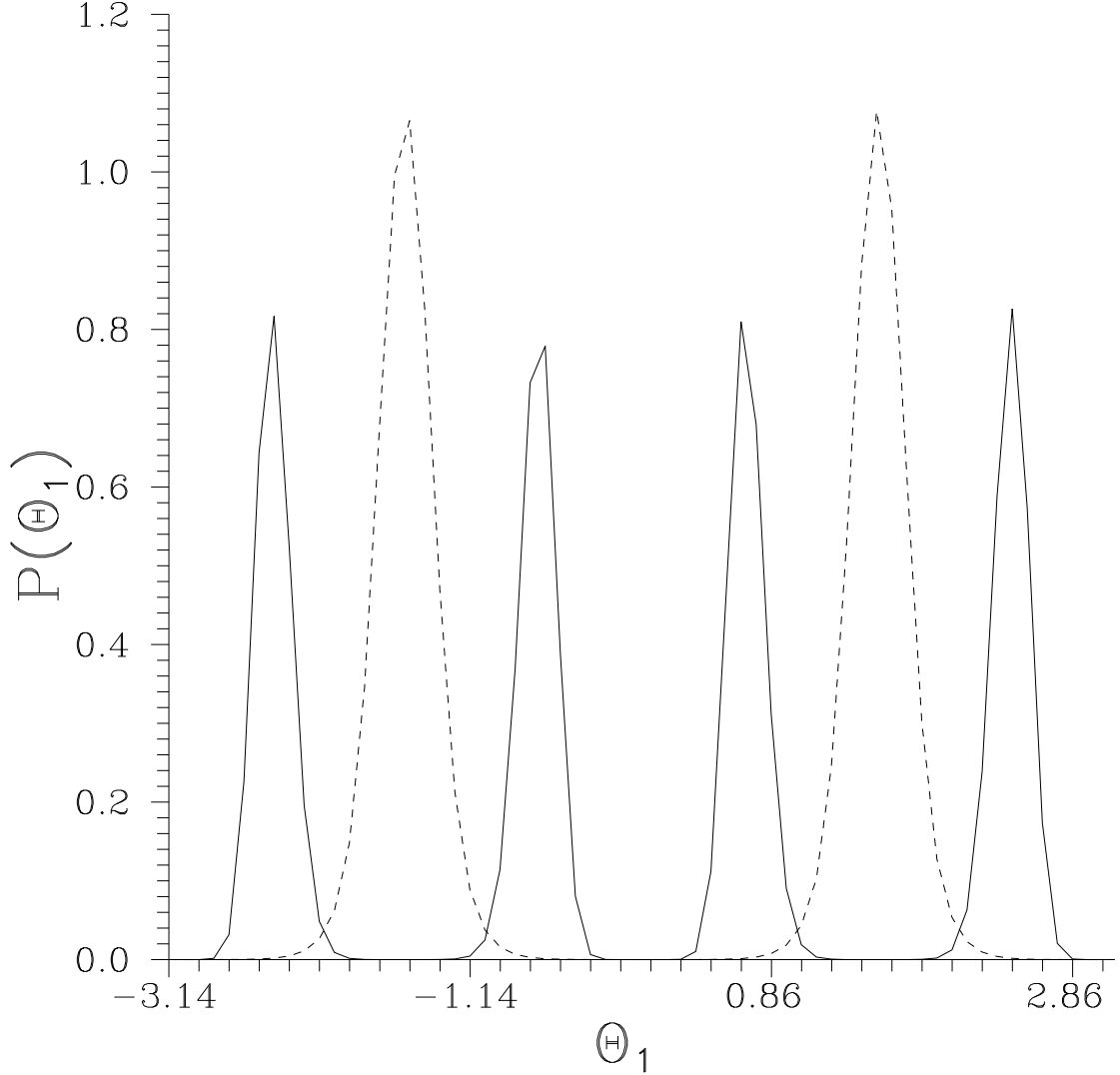}
\caption{ The phase distribution $P(\Theta_{1})$  of the
single-mode JCM, i.e. $k_{1}=0, k_{2}=1$, when $|\alpha_{1}|=5,
\varphi=0,\phi=0, \epsilon_{1}=1$ and for
  $T=7$ (solid curve-collapse time) and $16.4$ (dashed curve--
revival time).}
\end{figure}
 \begin{equation}
  \lambda_{1}=2T\sqrt{n(m+1)},\\
  \\
  \lambda_{2}=2T'\sqrt{2n(2m+1)}.
  \end{equation}
The revivals in $\langle \hat{\sigma}_{z}(T)\rangle$ for both cases occur
  when
  $\lambda_{1}\simeq 2m'\pi,\quad \lambda_{2}\simeq 2m'\pi$, where $m'$
  is a positive integer.
    In the strong-intensity regime
    $\lambda_{1}$ and $\lambda_{2}$
 can be simplified  as
  \begin{equation}
  \lambda_{1}=2T\sqrt{\bar{n}\bar{m}},\\
  \lambda_{2}=4T'\sqrt{\bar{n}\bar{m}}.
  \end{equation}
Therefore, the relation between the revival times of the two cases
is $T'_{R}=\frac{T_{R}}{2}$, where the subscript $R$ indicates that
this time is associated with the revival pattern.
On the other hand, in Fig. 1(b) the curve $A$ is given for $k_{j}=1$ when one of the
modes is initially in coherent states and the other is in even
coherent states; while the curve B is given for $k_{j}=2$ and the
modes are initially in even coherent states. The secondary revivals in
curve A are more pronounced than those in Fig. 1(a). Nevertheless, the curve B,
 which is given for
higher-order photon transition, exhibits chaotic behavior resulting from
the entanglement between different modes. This can be realised by comparing
this behavior with that of
 the atomic inversion of the two-photon-single-mode JCM
\cite{faisal1}, which exhibits compact and systematic revivals.
From the above discussion we can conclude that the occurrence of the RCP and
the secondary revivals in $\langle \hat{\sigma}_{z}(T)\rangle$ for the TMJCM
basically depends  on the values of the
transition parameters $k_{j}$ and on the type of the initial distribution
of the modes. Nevertheless, it
 is insensitive to the values of the initial intensities of the modes.

Now we draw the attention to the behavior of the $W$ function, in
particular, the evolution of the $W$ function at the phase space origin.
We start the discussion by assuming that the two modes are initially
prepared in even coherent states and $k_{1}=k_{2}=$odd number.
In this case (\ref{19a})--(\ref{19c}) reduce to
\begin{equation}
W_{1}(0,T)=\frac{1}{\pi}\langle\hat{\sigma}_{z}(T)\rangle,\quad
W_{2}(0,T)=\frac{-1}{\pi}\langle\hat{\sigma}_{z}(T)\rangle,\quad
W(0,T)=\frac{-1}{\pi^{2}}. \label{extra2}
\end{equation}
Expressions (\ref{extra2}) provide an important fact:
The evolution of the phase-space origin of the single-mode $W$ function
can carry information on the atomic inversion of the system.
However, the joint $W$ function is localized in phase space
giving maximum nonclassicality, i.e. maximum negativity.
On the other hand, when the number of photons involved in the atomic
transition is even, i.e. $k_{1}=k_{2}=$even number, the
 $W$ functions of  the single mode and
compound modes are localized in phase space:
\begin{equation}
W_{1}(0,T)=
W_{2}(0,T)=\frac{1}{\pi},\quad
W(0,T)=\frac{1}{\pi^{2}}. \label{extra3}
\end{equation}
Moreover, when $k_{1}$ is odd, say, and $k_{2}$ is even we obtain
\begin{equation}
W(0,T)=\frac{1}{\pi}W_{1}(0,T)=\frac{1}{\pi^{2}}
\langle\hat{\sigma}_{z}(T)\rangle,\quad
W_{2}(0,T)=\frac{1}{\pi}. \label{extra4}
\end{equation}
Finally, when $k_{1}, k_{2}$ are odd and the two modes are initially prepared
 in a general quantum states, e.g. coherent states--restricting the
 discussion to the excited atomic state--(\ref{19a}) and (\ref{19c}) take the forms
\begin{equation}
W_{1}(0,T)=\frac{1}{\pi}\sum_{n,m=0}^{\infty}C^{2}_{n,m+k_{2}}\cos(2T\Lambda_{n,m}+n\pi),
\label{extra4}
\end{equation}
\begin{equation}
W(0,T)=- W_{1}(0,0)W_{2}(0,0), \label{extra5}
\end{equation}
where $W_{j}(0,0)$ is the initial value of the $W$ function of the $j$th mode
at the phase space origin. Expression (\ref{extra4})--apart from the
prefactor $1/\pi$--is typical to that of the
$\langle\hat{\sigma}_{z}(T)\rangle$ but with additional factor, i.e.
$(-1)^n$, however,
the joint $W$ function is localized in the phase space.
 Actually, we have numerically checked the behavior of
(\ref{extra4}) and found that it can provide typical curves to those of
the corresponding atomic inversion.
 Therefore, for specific values of the interaction parameters the
evolution of the $W$ functions at the phase space origin represents the
atomic inversion of the system, in particular, when $k_{1}$ or $k_{2}$
(or both of them) is an odd number.
\begin{figure}
   \includegraphics[width=.90\linewidth]{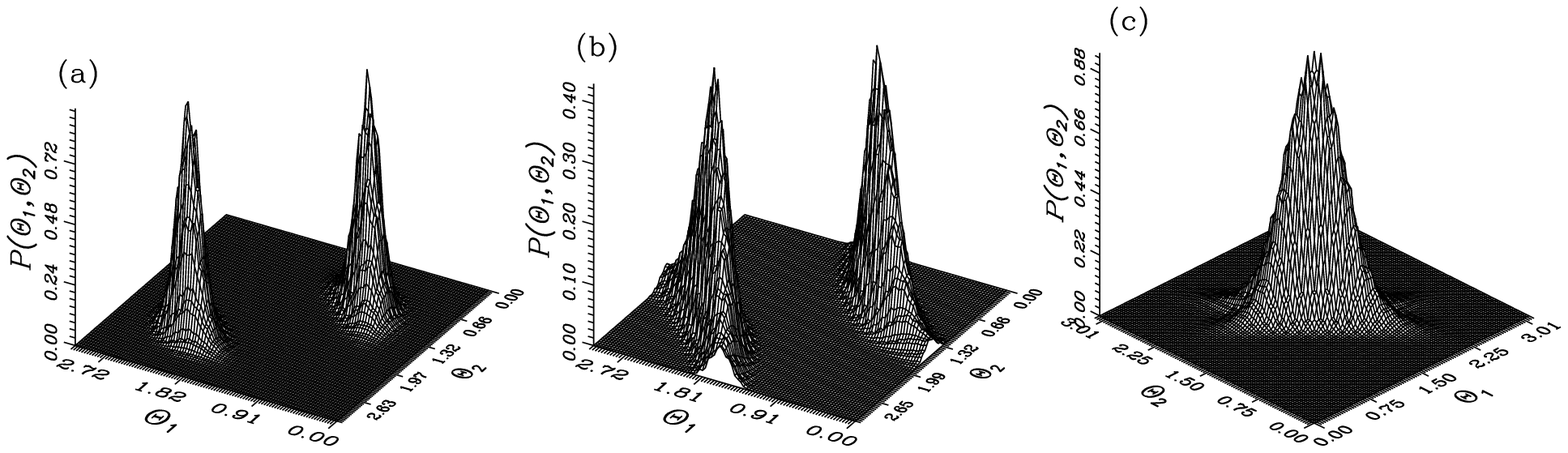}
\caption{ The joint phase distribution  $P(\Theta_1,\Theta_2)$ for
$|\alpha_{j}|=5,\varphi=0,\phi=0,k_{1} =k_{2}=1,
(\epsilon_{1},\epsilon_{2})=(1,1)$ and for
  $T=1.8$ (a), $4.4$ (b) and $3.22$ (c).}
\end{figure}
This leads to the fact that for TMJCM there
is a connection between the behavior of both the atomic inversion,
the $W$ function and the phase space distribution of the single mode
as well as the compound modes, as we shall show shortly.
This situation is similar to that of the standard JCM \cite{phas1}.

\subsection{Phase distribution }
For the standard JCM  it has been shown that there is a
relationship between the time evolution of
the atomic inversion,  the phase distribution and $Q$
function. For instance, the RCP is reflected in
the behavior of the phase distribution of the radiation field \cite{phas1} where the
single-peak structure of the initial
phase distribution splits into two peaks, which are rotated in opposite
directions. At this moment $\langle\hat{\sigma}_{z}(T)\rangle$ exhibits
collapse pattern. Furthermore, as the interaction proceeds the
 two peaks of the  distribution are collided showing
single-peak form, however, $\langle\hat{\sigma}_{z}(T)\rangle$ shows
revivals.  Here we show that similar behavior can occur
for the single-mode and  compound-mode cases of the TMJCM.
Additionally, we investigate the influence
of the higher-order photon transition $k_{j}>1$ on such behavior.
It is worth
reminding that
the behavior of the $\langle\hat{\sigma}_{z}(T)\rangle$ of the
TMJCM is rather complicated compared to that of the standard JCM (see Figs. 1).
\begin{figure}
   \includegraphics[width=.9\linewidth]{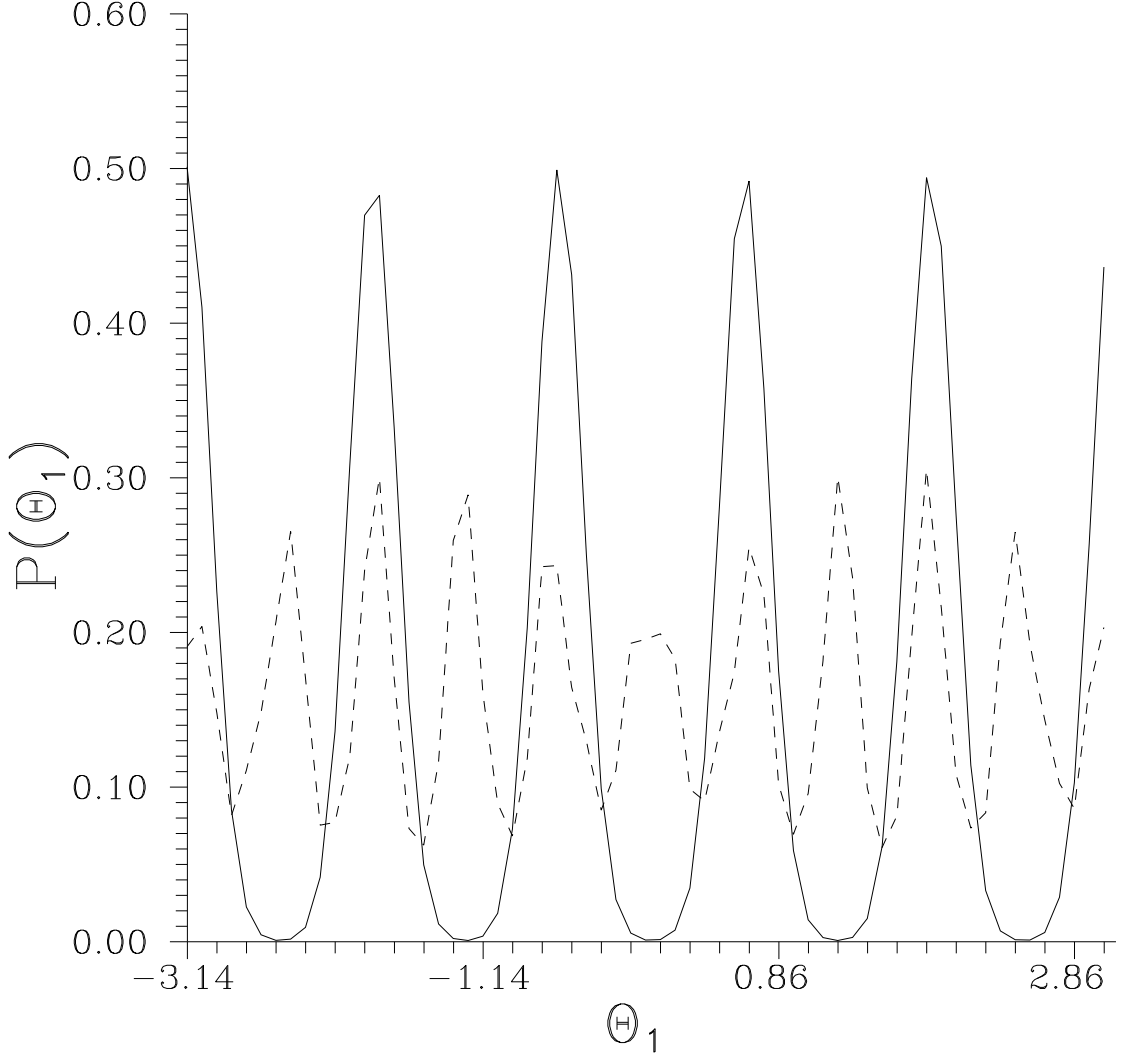}
\caption{ The phase distribution  $P(\Theta_{1})$ of the first
mode for $T=5,
 |\alpha_{j}|=5,\varphi=0,\phi=0,k_{1}=k_{2}=2$ and for $(\epsilon_{1},
 \epsilon_{2})=(0,0)$ (solid curve)  and  $(1,1)$ (dashed curve).
}
\end{figure}
Actually, we found that the behavior of the first and second modes
phase distribution are typical so that the discussion would be restricted to
that of the first mode only.
Furthermore, as it is difficult to  treat the behavior of the phase
distributions analytically
we use the numerical technique  (see Figs. 2--5).
Figs. 2(a) and (b) are given for the single-mode phase distribution
 when the two modes
are initially prepared in coherent and even coherent states,
respectively, for collapse time (solid curve), secondary revival time
(long-dashed curve) and revival time (short-dashed curve). In Fig. 2(b)
the star-centered curve is given for the case $(\epsilon_{1},\epsilon_{2})=(0,1)$
and the corresponding revival time $T=6.3$. In Fig. 2(a)
one can see that $P(\Theta_{1})$ exhibits two-peak structure for
both collapse time (with two maxima around $\Theta_{1}\simeq\pm 2\pi/3)$ and
secondary-revival time (with two maxima  around $\Theta_{1}\simeq \pm \pi/2)$,
 however, those of the latter
are border and shorter than those of the former.
In fact such behavior (, i.e., the occurrence of the two-peak structure in the phase
distribution) is quite similar to that of the even (odd) coherent
states \cite{buz1}. Furthermore, this behavior  is representative
 to the standard JCM indicating generation of the cat
states.  Nevertheless, it is difficult to consider  such conclusion here
owing to the
strong entanglement between different components of the system.  For the
revival time $P(\Theta_1)$ exhibits two wings around
$\Theta_{1}\rightarrow \pm\pi$ (see short-dashed
curve), i.e. it is incomplete peak. In Fig. 2(b),
 one can see that four-peak structure occurring through the collapse and
secondary revival times for the initial even coherent states, which are
combined at the revival time  providing two-peak structure.
Interestingly, when one of the modes is initially prepared in coherent
state and the other is in even coherent state we have seen that for collapse and
secondary revival times the behavior is typical to that of the two modes, which are
initially prepared in the cat states (see solid and long-dashed curves in
Fig. 2(b)), however, for the revival time we obtain single-peak around
$\Theta_{1}\simeq 0$ and two wings as $\Theta_{1}\rightarrow \pm \pi$
(see the star-centered curve in Fig. 2(b)). This indicates that the
influence of the interference in phase space on the behavior of the phase
distribution is more pronounced in the course of the
revival time. Nevertheless, the influence of the entanglement on the behavior of
the distribution can be realised by
comparing Fig. 3, which is given for the single-mode JCM when the mode
is initially prepared in the cat state, with the Fig. 2(b), in
particular, the solid and short-dashed curves in the two figures. Such
comparison shows that there is
agreement and disagreement in the phase distribution of the two systems. For instance,
 the two systems exhibit four (two)
peaks in the course of the collapse (revival) time, however,
 those of the single-mode JCM are narrower and
higher  than those of the TMJCM.  The reason for occurring the same number
of peaks in the distribution of the two systems is the term
$\cos((n-n')\Theta_{1})$.
Also from (\ref{extra2}) and (\ref{extra4}) Figs. 2 can
 provide information on the evolution of the $W$ function at the
phase space origin.

Now we turn the attention to the joint phase distribution.
We have noted that the evolution of the
 $P(\Theta_{1},\Theta_{2},T)$ reflects well the behavior of the
$\langle\hat{\sigma}_{z}(T)\rangle$ (see Figs. 4(a)--(c) for given values
of the parameters, which are the same as those of Fig. 2b). Based on the fact  $P(\Theta_{1},\Theta_{2},T)$
is symmetric in $\Theta_{1}$ and $\Theta_{2}$ we have plotted Figs. 4 only
over the region $0\leq \Theta_{j}\leq \pi$.
Generally, there is a similarity between the single-mode and compound-mode
phase distribution. To be more specific,
 for initial
even coherent light, $P(\Theta_{1},\Theta_{2},T)$ exhibits two peaks
over the specified region in
the course of the collapse and secondary revival times, which reduce to single peak
(recombination of the peaks) through the revival time. Moreover, the peaks, which
are related to the secondary revival time, are broader than those of the
collapse time (compare Fig. 4(a) and Fig. 4(b)).
Also the influence of the interference in phase space
on the behavior of the  joint phase
distribution is remarkable as multipeak structure.

We conclude this part by shedding the light on the two facts.
(i) For $k_{j}>1$ where the atomic inversion
exhibiting chaotic behavior the phase distributions show multipeak
structure, which is insensitive to the values of the interaction time
(, i.e. steady state phase distribution).
Of course the number of peaks for the initial cat states are approximately
two times greater than those  for the initial coherent light (see Fig. 5 for given
values of the parameters).
(ii) As it is well known that for the standard JCM "coherent trapping"
(, i.e. $\langle\hat{\sigma}_{z}(T)\rangle\simeq 0$)
can occur \cite{zaheer}.
Similarly this can occur for the TMJCM under certain condition
(cf. (\ref{10a})).
Nevertheless, this cannot be remarked in the behavior of the phase
distribution as a result of the fact that the phase distribution is
normalized (cf. (\ref{12})).
Therefore, the relationship between
the behavior of $\langle\hat{\sigma}_{z}(T)\rangle, P(\Theta_{j},T)$
and $P(\Theta_{1},\Theta_{2},T)$ occurs only when  the two-level atom is
either in the excited or ground state. Thus we have
limited the discussion
to these cases. Finally, we have found also that the behavior
of $P(\Theta_{j},T)$ and $P(\Theta_{1},\Theta_{2},T)$ are independent
of the type of the initial cat states and the initial atomic states.

\begin{figure}
   \includegraphics[width=.9\linewidth]{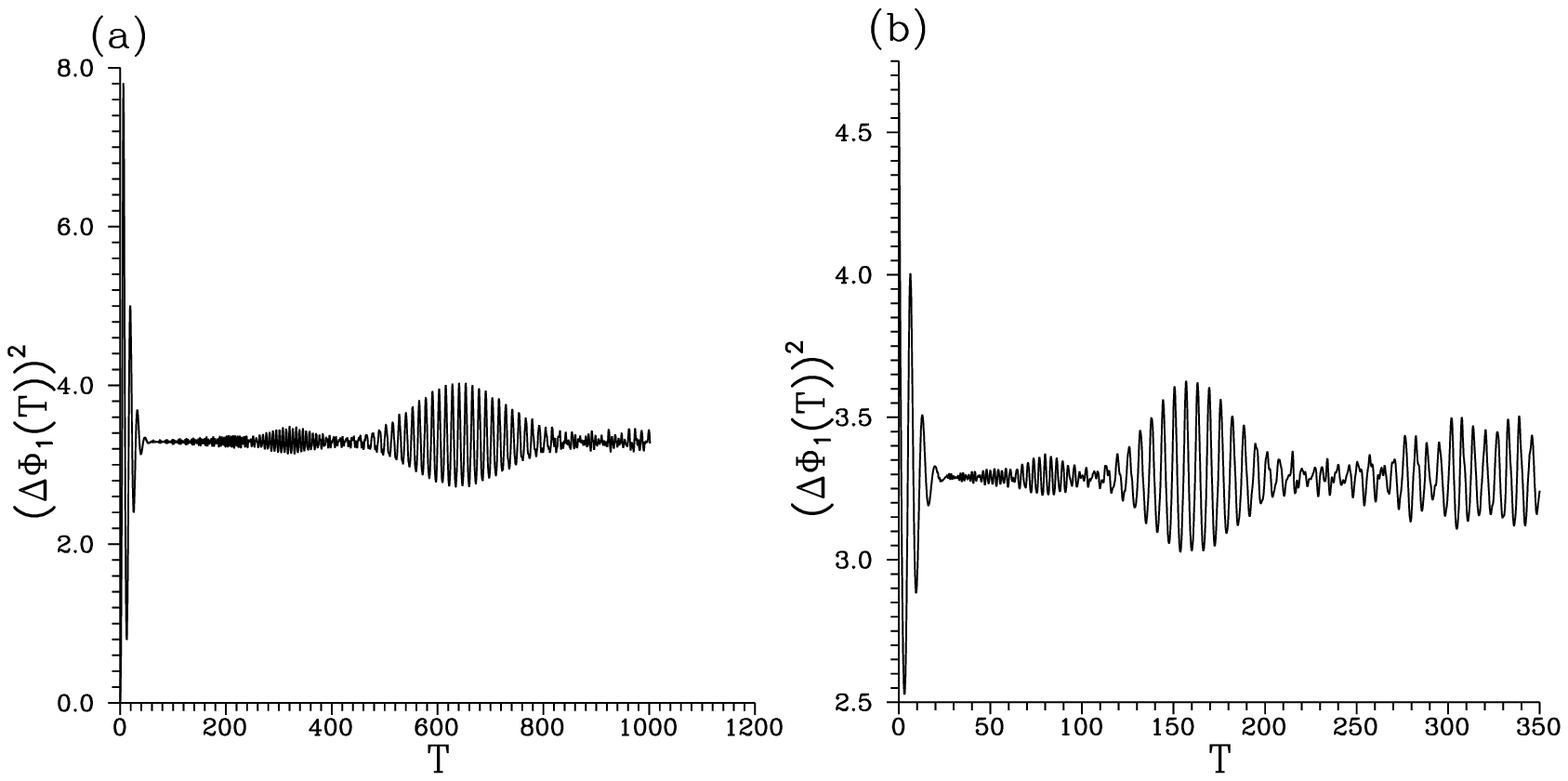}
\caption{ The single-mode phase variance $\langle
(\triangle\hat{\Phi}_{1})^{2}\rangle$ of the first mode versus the
scaled time $T$ for $|\alpha_{j}|=5,\varphi=0,\phi=0,
k_{1}=k_{2}=1$, and for $(\epsilon_{1},\epsilon_{2})=(0,0)$ (a)
and $(1,1)$ (b). }
\end{figure}

\subsection{Phase fluctuations}
For the standard JCM it has been shown that the phase
variance exhibits collapses and revivals about the long-time interaction, and
there appear main revivals and secondary revivals with different periods
\cite{phas2}. The origin of this phenomenon is in the phase correlations
between the different  eigenstates of the photon number that make
contributions to the orientated distribution of phase, and the
contribution due to the correlations between the neighboring Fock states
is much larger than the others \cite{phas2}. Actually, we have seen that this
phenomenon can occur
for the TMJCM for interaction time  shorter than that of the
single-mode JCM (see Figs. 6--7 for given values of the interaction
parameters). From Fig. 6(a) we observe that the RCP as well as
secondary revival occur around  $\pi^{2}/3$, around the random
phase distribution. Comparison of this figure with the Fig. 1 in
 \cite{phas2} shows that the entanglement between the two modes
in the interaction with the two-level atom makes
the RCP in the phase variances occurring through interaction time
several times smaller than that of the standard JCM.
 This behavior can be roughly explained in the following sense. We
 restrict the analysis to the case $k_{1}=k_{1}=1$ when the two modes
are initially prepared in the coherent states with strong intensities
 (, i.e. $|\alpha_{j}|
 >>1, \quad j=1,2$) and the atom is in the excited state.
From (\ref{15a}) the phase variance of the first mode
 $\langle (\triangle\hat{\Phi}_{1})^{2}\rangle$ takes the form

\begin{eqnarray}
\begin{array}{lr}
\langle
(\triangle \hat{\Phi}_{1})^{2}\rangle=\frac{\pi^{2}}{3}
+4\sum\limits_{n> n'}^{\infty}
\sum\limits_{m=0}^{\infty}
C_{n}^{(1)}
C_{n'}^{(1)}
(C_{m+1}^{(2)})^{2}\frac{(-1)^{n'-n}}{(n'-n)^{2}}
\\
\\
\times
\cos[T\sqrt{m+1}(\sqrt{n+1}-\sqrt{n'+1})].
\label{155a}
\end{array}
\end{eqnarray}
It is worth mentioning that the expression of the
phase variance of the standard JCM can
be obtained from (\ref{155a}) by dropping
the summation $\sum_{m=0}^{\infty}$ and the notations related to the
second mode. In this case
the argument of the $\cos(.)$, i.e. the argument of the
dynamical term of the standard JCM, is
\begin{equation}
TZ_{n,n'}=T(\sqrt{n+1}-\sqrt{n'+1}). \label{155b}
\end{equation}
Therefore, the phase variance
 can exhibit revivals, i.e. $\cos(.)$ provides maxima, only when
\begin{equation}
TZ_{n,n'}=l\pi, \quad l \;\;\; is\;\; integer \label{155c}
\end{equation}
where $T$ is the revival time, say, of the phase variance of the
standard JCM.
Now we draw the attention to the $\langle
(\triangle \hat{\Phi}_{1})^{2}\rangle$ of the TMJCM
given by (\ref{155a}).
 In the strong-intensity regime, i.e. $\sqrt{\bar{m}}=|\alpha_{2}|
>>1$,
  we can apply the harmonic approximation technique
\cite{faisal1} to evaluate the summation related to the second mode
where $(C_{m}^{(2)})^{2}$ has Poissonian
distribution.
In this case (\ref{155a}) reduces to
\begin{eqnarray}
\begin{array}{lr}
\langle(\triangle \hat{\Phi}_{1})^{2}\rangle=\frac{\pi^{2}}{3}
+4\sum\limits_{n> n'}^{\infty}
C_{n}^{(1)}
C_{n'}^{(1)}
\frac{(-1)^{n'-n}}{(n'-n)^{2}}
\\
\\
\times
\exp[-2\bar{m}\sin^{2}(\frac{TZ_{n,n'}}{4\sqrt{\bar{m}} })]
\cos[\frac{1}{2}T\sqrt{\bar{m}}Z_{n,n'}+\bar{m}\sin(\frac{TZ_{n,n'}}
{4\sqrt{\bar{m}} })
].
\label{155d}
\end{array}
\end{eqnarray}
Expression (\ref{155d}) can provide revivals when the exponential term is
maximum, i.e.
\begin{equation}
\frac{T'Z_{n,n'}}{4\sqrt{\bar{m}}}=l\pi, \label{155g}
\end{equation}
where $T'$ denotes the revival time of the $\langle(\triangle \hat{\Phi}_{1})^{2}\rangle$
of the TMJCM.
From (\ref{155c}) and (\ref{155g}) we arrive at
\begin{equation}
\frac{T'}{4\sqrt{\bar{m}}}=T. \label{155h}
\end{equation}
Expression (\ref{155h}) shows that $T'<T$. In other words, the RCP in
$\langle(\triangle \hat{\Phi}_{1})^{2}\rangle$ for the TMJCM occurs
for interaction time $4\sqrt{\bar{m}}$ times shorter than that of the
standard JCM.
We proceed that the RCP
in $\langle(\triangle \hat{\Phi}_{1})^{2}\rangle$ can occur through
interaction time smaller than that shown in Fig. 6(a) by preparing
 the two modes initially  in the cat states.
This is shown in Fig. 6(b) and can be proved "roughly" using procedures
 as those given above. Moreover, the explanation used for the behavior of the
 atomic inversion of this case can be adopted here.

On the other hand, for the evolution of the phase variances of the standard JCM there
are two important cases, which are three- and four-photon transition,
i.e. $k_{1}=0$, $k_{2}=3$ and $4$.  The phase variances for these cases exhibit
significant RCP, in particular, the four-photon transition  case
whose phase variance exhibits periodic  revivals  with  period $\pi$.
Moreover, these revivals  almost restore their initial amplitudes and
widths, and they are independent of the intensities of the initial modes
\cite{phas2}. Surprisingly, we have noted--apart from the
amplitudes of the revivals--that there is a
one-to-one correspondence between the behavior of the phase variance
of the standard  JCM
and that of the TMJCM.  In other
words, we have seen that the behavior of the single-mode phase variances
of the  cases  $k_{1}+k_{2}=3$
(provided that $k_{j}\neq 0,\quad j=1,2$) and
$k_{1}=k_{2}=2$ associated with the TMJCM are similar
 to those of the standard JCM.   This information can  be realised from Figs. 7(a) and
(b). From Fig. 7(a) we can see a systematic behavior, i.e. each revival
is followed by secondary revival and the initial revivals are
restored.  Furthermore, the revivals and secondary revivals  are
intensities independent (provided that $|\alpha_{j}|>>1$) and they occur
periodically with period $\pi/2$, i.e. half of that of the standard JCM.
The origin of such  behavior is in the  two-photon transition and
 in the  interference in phase space
(compare Fig. 7(a) with Fig. 4 in \cite{phas2}). Furthermore, the comparison
of Fig. 7(b) with Fig. 3 in \cite{phas2} is instructive, where one can
realise the role of the entanglement and the interference in phase space.

\begin{figure}
  \includegraphics[width=.9\linewidth]{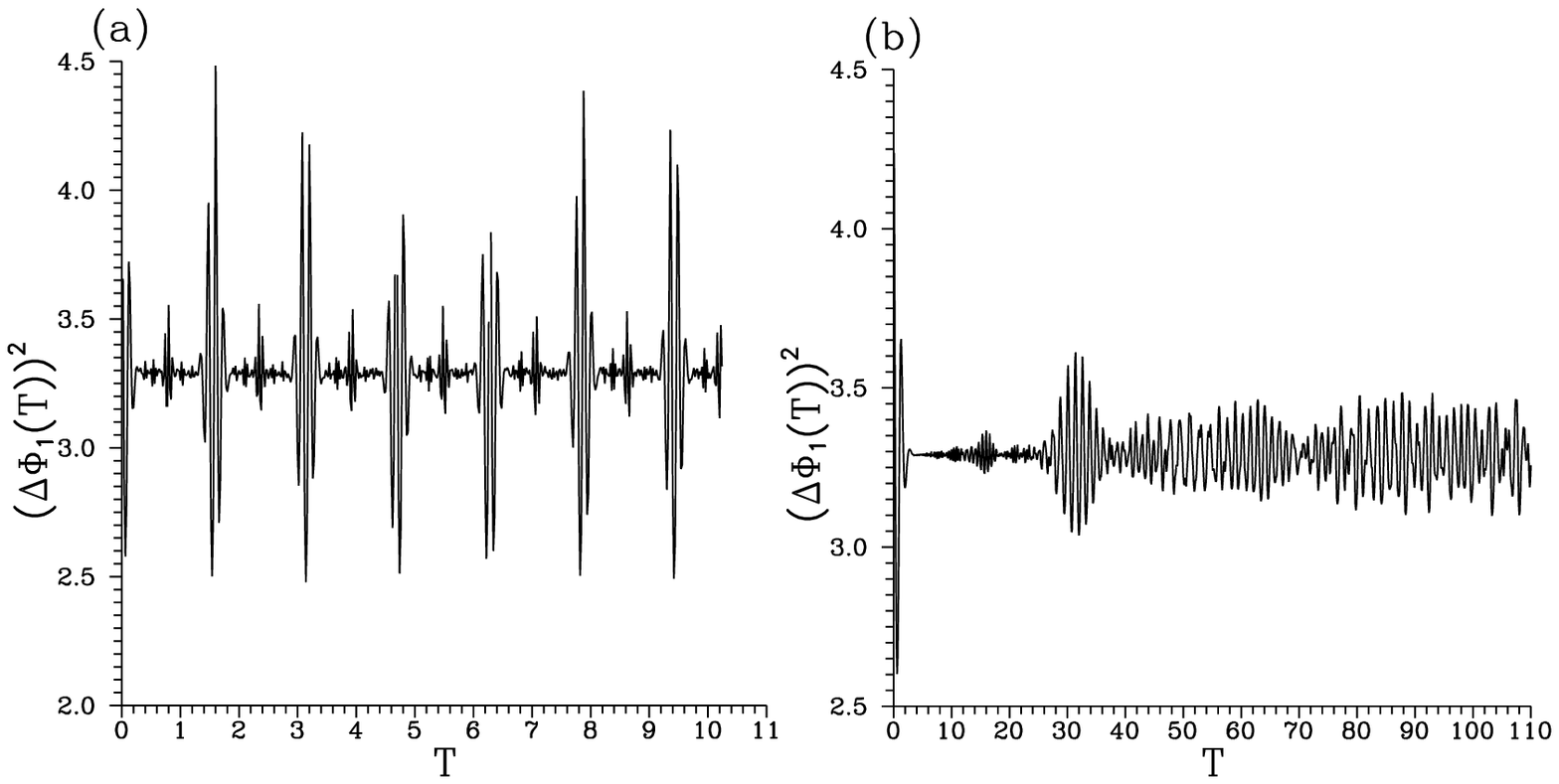}
\caption{ The single-mode phase variance $\langle
(\triangle\hat{\Phi}_{1})^{2}\rangle$ of the first mode versus the
scaled time $T$ for
$|\alpha_{j}|=5,\epsilon_{j}=1,\varphi=0,\phi=0$,
$(k_{1},k_{2})=(2,2)$ (a) and $(1,2)$ (b). }
\end{figure}

Generally, the occurrence of the RCP in the $\langle (\triangle\hat{\Phi}_{1})^{2}\rangle$
cannot occur in the
$\langle (\triangle\hat{\Phi}_{\pm})^{2}\rangle$
where  $h_{1,2}\neq 0$
(cf. (\ref{15a})).
 We  shed
the light on the behavior of the mean-photon number variances.
It is worth reminding that the mean-photon numbers and the phase operators
are conjugate quantities (noncommuting operators). In Figs. 8(a), (b) and (c)
we plot $\langle (\triangle \hat{n}_{1})^{2})\rangle,
\langle (\triangle \hat{n}_{+})^{2})\rangle$ and
$\langle (\triangle\hat{n}_{-})^{2}\rangle$, respectively,
for the given values of the interaction parameters.  It is
obvious that
$\langle (\triangle \hat{n}_{1})^{2})\rangle$ and
$\langle (\triangle\hat{n}_{-})^{2}\rangle$ exhibit RCP (see Fig. 8(a) and (c))
which is insensitive to the intensities of the initial fields.
Moreover, this behavior  is almost similar to that of the corresponding atomic
inversion. In Fig. 8(b) we can see that the chaotic behavior is dominant.
To be more specific, the behavior
of $\langle (\triangle\hat{n}_{+})^{2}\rangle$
 is close to the steady state,
$\langle (\triangle \hat{n}_{+}(T))^{2})\rangle\simeq
\langle (\triangle \hat{n}_{+}(0))^{2})\rangle$.
Comparison between Fig. 8(b) and Fig. 8(c) shows that the correlation term,
i.e. $\langle \hat{n}_{1}\hat{n}_{2}\rangle
-\langle\hat{n}_{1}\rangle\langle \hat{n}_{2}\rangle$,
plays a constructive (destructive) role in the behavior of
the $\langle (\triangle\hat{n}_{-})^{2}\rangle$
($\langle (\triangle\hat{n}_{+})^{2}\rangle$).

\begin{figure}
   \includegraphics[width=.9\linewidth]{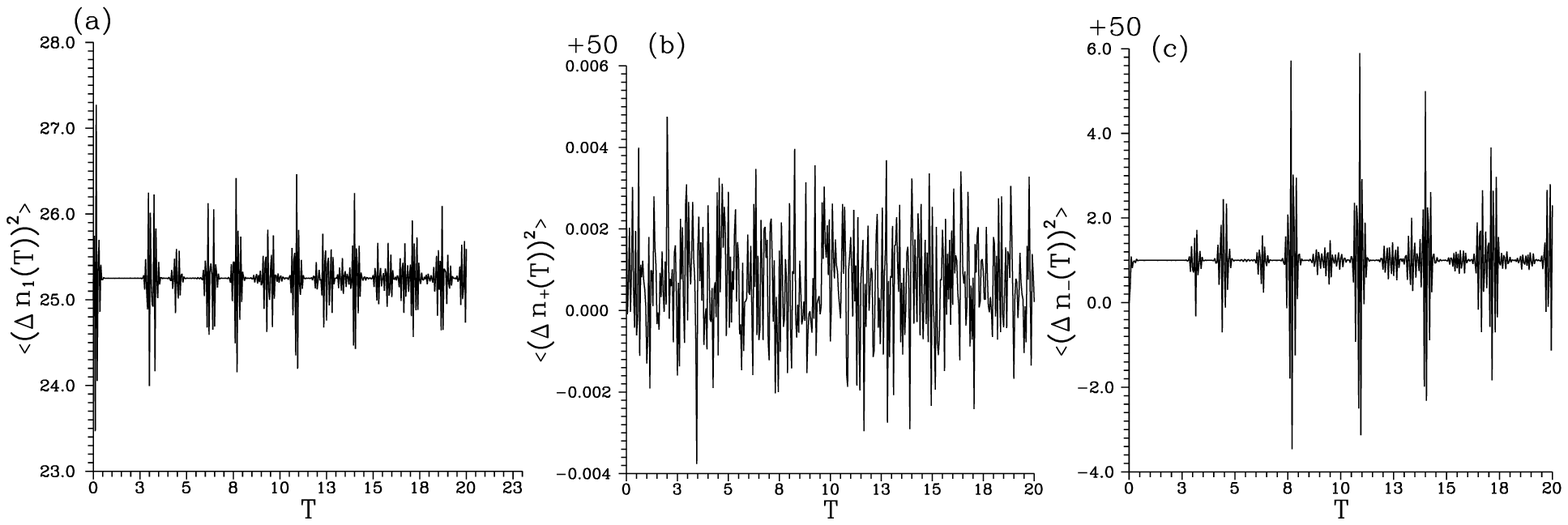}
\caption{ The mean-photon number variances versus the scaled time
$T$
 for  $|\alpha_{j}|=5,\epsilon_{j}=1,\varphi=0,\phi=0,
(k_{1},k_{2})=(1,1)$ and for
the single-mode case (a), sum-mode case (b) and difference-mode case
(c).
}
\end{figure}
Now we can conclude that
$\langle (\triangle\hat{\Phi}_{1})^{2}\rangle$  can
provide  RCP,  which is insensitive to the type of initial atomic
state. We have to stress that the
RCP in the atomic  inversion and
in the phase variances are completely different and cannot be compared.
Finally, it is worth mentioning that RCP has
 been realised  also for the   phase variances of the
  two-photon down-conversion with quantized pump, e.g. \cite{gan}.

\section{Conclusions}
In this paper we have discussed the phase properties
of the lossless multiphoton two-mode Jaynes-Cummings model
for Schr\"{o}dinger-cat states in the framework
of Pegg-Barnett formalism.
The investigation has included also the
dynamical behavior of the $W$ function at the phase space origin. We
have shown that the evolution of the
$\langle\hat{\sigma}_{z}(T)\rangle$ of the TMJCM is reflected in the
behavior of the phase distribution
where the splittings and overlappings of the phase distribution are
accompanied by the collapses and revivals, respectively, in $\langle
\hat{\sigma}_{z}(T)\rangle$. This has been remarked in the behavior
of both the single-mode
and compound-mode phase distributions. Furthermore, in the course of the
secondary revival in $\langle \hat{\sigma}_{z}(T)\rangle$ the phase
distribution exhibits behavior almost similar to that of the collapse case.
Moreover, the nonclassical multipeak structure, which is representative to
cat states, is remarked in the behavior of the phase distribution.
 Nevertheless, this behavior  is insensitive to
the type of both the initial cat state (, i.e. if the modes are initially
prepared in even or odd coherent
states) and the initial atomic states (, i.e. if it is in the excited or
ground state). Additionally, we have shown that for higher-order photon
transition, i.e. $k_{j}>1$, the phase distribution is in the
steady state showing
multipeak structure.  Also we have proved that under certain condition
there is a clear relationship between the atomic inversion and $W$
function. Such relation leads to the fact: There is a
relationship between the $W(0,T)$ and the corresponding
phase distribution.
Finally, we have shown that the single-mode phase variances can exhibit
RCP about the long-time interaction.

\section*{Acknowledgement}
One of the authors (F.A.A.E.) is grateful to the International
Islamic University Malaysia for hospitality and financial support.
\section*{References}

\end{document}